%% file: main.tex
\documentclass[10pt, conference, letterpaper]{IEEEtran}
\IEEEoverridecommandlockouts
\usepackage{float}  
\usepackage{cite}
\usepackage{amsmath,amssymb,amsfonts}
\usepackage{amsmath}  
\usepackage{subcaption}
\usepackage{graphicx}
\usepackage{booktabs}
\usepackage{textcomp}
\usepackage{multirow}
\usepackage{xcolor}
\usepackage{xspace}
\usepackage{algorithm}
\usepackage{algpseudocode}

\def\BibTeX{{\rm B\kern-.05em{\sc i\kern-.025em b}\kern-.08em
    T\kern-.1667em\lower.7ex\hbox{E}\kern-.125emX}}
\begin{document}

\title{\vspace{-1em} SmartFLow: A Communication-Efficient SDN Framework for Cross-Silo Federated Learning \\
\vspace{0.2em}
\footnotesize This work has been submitted to the IEEE for possible publication.\\ Copyright may be transferred without notice, after which this version may no longer be accessible.
}

\author{
\IEEEauthorblockN{Osama Abu Hamdan, Hao Che}
\IEEEauthorblockA{
\textit{University of Texas at Arlington}\\
oma8085@mavs.uta.edu, hche@cse.uta.edu}
\vspace{-3mm}
\and
\IEEEauthorblockN{Engin Arslan}
\IEEEauthorblockA{\textit{Meta} \\
enginarslan@meta.com}
\vspace{-3mm}
\and
\IEEEauthorblockN{Md Arifuzzaman}
\IEEEauthorblockA{
\textit{Missouri University of Science and Technology}\\
marifuzzaman@mst.edu}
\vspace{-3mm}
}

\maketitle

\newcommand{\name}{\textit{SmartFLow}\xspace}
\newcommand{\greedy}{$SmartFLow_{Greedy}$\xspace}
\newcommand{\cp}{$SmartFLow_{CP}$\xspace}
\newcommand{\rfwd}{\textit{RFWD}\xspace}
\newcommand{\freecap}{\textit{FreeCap}\xspace}
\newcommand{\openflow}{\textit{OpenFlow}\xspace}

\begin{abstract}
Cross-silo Federated Learning (FL) enables multiple institutions to collaboratively train machine learning models while preserving data privacy. In such settings, clients repeatedly exchange model weights with a central server, making the overall training time highly sensitive to network performance. However, conventional routing methods often fail to prevent congestion, leading to increased communication latency and prolonged training. Software-Defined Networking (SDN), which provides centralized and programmable control over network resources, offers a promising way to address this limitation. To this end, we propose \name, an SDN-based framework designed to enhance communication efficiency in cross-silo FL. \name dynamically adjusts routing paths in response to changing network conditions, thereby reducing congestion and improving synchronization efficiency. Experimental results show that \name decreases parameter synchronization time by up to $47\%$ compared to shortest-path routing and $41\%$ compared to capacity-aware routing. Furthermore, it achieves these gains with minimal computational overhead and scales effectively to networks of up to $50$ clients, demonstrating its practicality for real-world FL deployments.
\end{abstract}


\section{Introduction and Motivation}
\input{introduction_motivation}

\section{Related Work}
\input{related}

\section{System Overview}
\input{design}

\section{Path Selection and Switching Strategies}\label{sec:path}
\input{path_select}

\section{Experimental Results}\label{sec:results}
\input{evaluations}

\section{conclusion}
\input{conclusion}

\bibliographystyle{IEEEtran}  
\footnotesize
\bibliography{ref}

\end{document}

%% file: introduction_motivation.tex
In modern distributed computing, communication efficiency has overtaken processing power as the main performance bottleneck \cite{intro6-zen}. Hardware improvements have reduced local computational delays in Machine Learning (ML), but network constraints remain a major challenge, especially in federated learning (FL), where clients collaboratively train a shared model without sharing raw data. Each client updates the model using private data and periodically sends parameters to a central server, which aggregates them into a global model. While FL enhances privacy, it incurs high communication costs due to frequent transmission of large parameter sets over often unreliable networks \cite{intro4-bonawitz, intro2-mcmahan}.

\begin{figure}[t]
    \centering
    \includegraphics[width=0.95\linewidth]{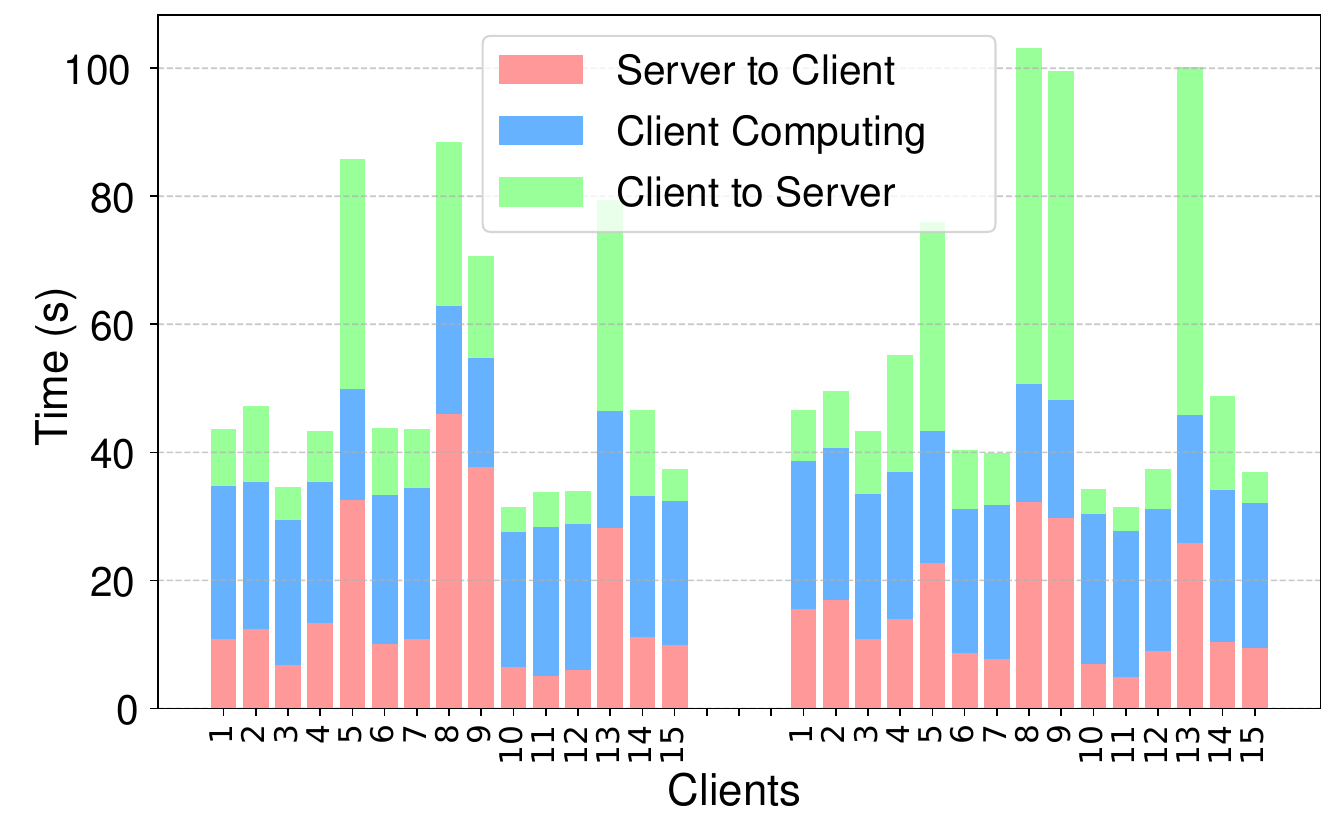}
    \vspace{-1mm}
    \caption{Communication dominates training time in cross-silo FL, with network congestion causing per-client delays, leading to synchronization bottlenecks.}
    \label{fig:cumulative_bar_plot}
    \vspace{-2mm}
\end{figure}

These issues intensify in cross-silo FL, where geographically distributed organizations rely on wide-area networks. Unlike centralized ML, which benefits from robust cloud infrastructure, cross-silo FL suffers from high latency, limited bandwidth, and routing inefficiencies that impair scalability \cite{intro1-kairouz, intro5-lyu}. As shown in \figurename~\ref{fig:cumulative_bar_plot}, training a MobileNet Large v3 model \cite{mobilenetlarge} across 15 clients on a synthetic Gabriel graph topology \cite{gabriel} revealed communication as the dominant factor in total training time. Uneven network conditions also delay some clients disproportionately, creating synchronization bottlenecks in synchronous FL. These delays reduce efficiency and slow convergence, underscoring the need to mitigate network-related constraints.

To address these network-centric challenges, Software-Defined Networking (SDN) offers a promising solution. SDN decouples network control from data forwarding, allowing centralized and programmable network management. By providing a global network view, SDN supports dynamic routing decisions, real-time monitoring, and proactive congestion control. These capabilities align closely with the centralized coordination inherent to FL, facilitating adaptive communication strategies tailored specifically to training requirements. SDN-based traffic engineering dynamically allocates network resources, effectively reducing latency, mitigating congestion, and efficiently utilizing bandwidth \cite{intro8-mendonca}.

In this paper, we propose \name, an SDN-integrated federated learning framework designed to enhance communication efficiency in cross-silo scenarios. By leveraging real-time network state information, \name optimizes client-server communication during training sessions. This approach reduces communication delays, alleviates network congestion, and mitigates synchronization issues commonly observed in FL deployments. Additionally, \name complements existing FL optimization techniques, such as parameter compression \cite{intro9-compression}, quantization \cite{intro8-quantization}, and selective updates \cite{intro10-dropping}, broadening its applicability across various FL settings. Experimental evaluations demonstrate that \name significantly reduces overall training time, achieving up to 47\% improvement compared to other routing approaches.

%% file: related.tex
Research on the integration of SDN and federated learning is still in its early stages. Ma et al. \cite{ma_survey} conducted a comprehensive survey outlining key challenges in SDN-based FL environments, such as incentive design, privacy concerns, and efficient model aggregation. Mahmod et al. \cite{mahmod_freecap} proposed an SDN-enabled approach tailored for time-sensitive FL applications, using dynamic network adaptation to reduce communication delays by mitigating overload conditions. Other efforts, such as those by Sabah et al. \cite{sabah_review}, focused on reducing communication overhead through methods like model compression, client selection, and asynchronous updates. Similarly, Konečný et al. \cite{konecny_communication_efficiency} introduced techniques such as structured updates and sketching to reduce uplink costs. However, these studies either omit SDN altogether or do not address network-layer inefficiencies like latency and congestion.

Our work complements and extends these contributions by emphasizing dynamic routing optimization as a central mechanism. While prior approaches mainly address communication at the protocol or model level, they do not fully exploit SDN’s capabilities for real-time network adaptation. By introducing SDN-enabled routing strategies, our framework targets latency reduction and improved scalability, directly addressing communication bottlenecks in cross-silo FL systems and filling a critical gap in the current literature.

%% file: design.tex
\begin{figure}[t]
	\begin{center}
		\includegraphics[width=0.90\linewidth]{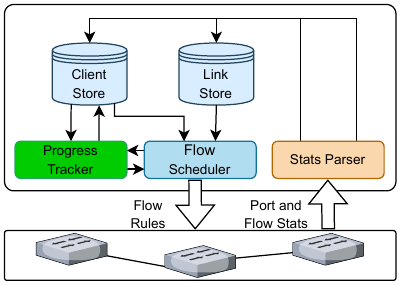} 
		\caption{\name implements three services to monitor the network, track training progress, and make routing decisions. It also utilizes two stores to keep track of link and flow statistics.}
		\label{fig:smartflow-design}
	\end{center}
    \vspace{-6mm}
\end{figure}

We implemented \name within the SDN ONOS controller \cite{onos}, using two data stores and three services to monitor network status and make routing decisions. The \emph{Stats Parser} processes \openflow statistics and updates the \emph{Client Store}, which holds dynamic, client-specific routing and traffic metrics, and the \emph{Link Store}, which captures performance and load conditions across network links. The \emph{Flow Scheduler} uses information from both stores to select efficient paths for each client and adjusts them as conditions change. It operates in coordination with the \emph{Progress Tracker}, which monitors data transfers and signals when updates complete. \figurename~\ref{fig:smartflow-design} shows the system architecture. The implementation is available as open-source at~\cite{smartflow-code}.

\subsection{Stats Parser} \label{renet-stats-parser}
The \emph{Stats Parser} periodically collects and analyzes \openflow statistics from network devices to optimize utilization and enhance active flows. Collected metrics include \textit{port statistics} for link congestion assessment and \textit{flow statistics} to estimate throughput per flow. This allows \name to track FL client and server progress in transmitting model weights relative to the model size. By default, the SDN server polls these statistics from each client's edge switch every five seconds.

This service includes a subcomponent that monitors one-way link delays and packet losses using Metter's method \cite{latency_and_loss}. It dispatches 10 Ethernet probe packets per link every second, each containing a link ID, sequence number, and timestamp. The controller computes one-way delays based on transmission, propagation, processing, and queuing times, using packet timestamps and arrival times. Packet loss rates, calculated over a sliding window of the last 100 probes, equal one minus the reception rate. These metrics are stored in the \emph{Link Store}, while the service runs efficiently in the background.

\subsection{Client and Link Stores} \label{client-link-stores}
The \emph{Client Store} maintains client-specific data, as summarized in \tablename~\ref{tab:flowstore_metrics}. Upon client joining, the system calculates the K-shortest server-to-client (S2C) and client-to-server (C2S) paths using ONOS's routing functionality, storing them for future use. The parameter $K$ is configurable based on network topology size. The store also tracks active paths adjusted by the \emph{Flow Scheduler}, monitors real-time throughput, and records data transfer volume per training round.

The \emph{Link Store} maintains detailed link statistics, including \emph{Default Capacity} and \emph{Current Throughput}. It identifies active client sets for C2S and S2C communications, informing potential rerouting impacts. Crucial metrics such as \emph{Packet Loss} and \emph{Link Latency}, affecting TCP efficiency \cite{tcp_rfc5681}, are tracked. To maintain stability despite measurement fluctuations, the store uses weighted averages of the three most recent latency and loss values.

\begin{table}[b]
    \caption{Information stored in \emph{Client Store} and \emph{Link Store} for client-flows and active-links characteristics}
    \vspace{-2mm}
	\renewcommand{\arraystretch}{1.2} 
	\begin{tabular}{l|l}
		\hline
		\multicolumn{2}{c|} { \textbf{Client Store Metrics}} \\ \hline
		\textit{S2C/C2S Paths}          & K-shortest paths between client and server    \\ 
		\textit{Current S2C/C2S Path}   & Current route of a S2C/C2S flow               \\ 
		\textit{Current Recv/Send Rate} & Current rate of a flow (Mbps)                 \\ 
		\textit{Round Sent/Recv Data}   & Downloaded or uploaded data / round (MB)      \\ \hline
		\multicolumn{2}{c|} { \textbf{Link Store Metrics}} \\ \hline
		\textit{Default Capacity}       & Base capacity of the link (Mbps)              \\
        \textit{Current Throughput}     & Instantaneous traffic rate on the link (Mbps) \\ 
		\textit{S2C/C2S Clients}            & Clients downloding/uploading data              \\
        \textit{Packet Loss}     &  Measured packet loss rate in the link \\ 
        \textit{Latency}     &   Measured one-way delay of the link (ms) \\ 
		\hline
	\end{tabular}
	\label{tab:flowstore_metrics}
\end{table}

\subsection{Flow Scheduler} \label{sec:flow_scheduler}
The \emph{Flow Scheduler} is a core component of \name, responsible for managing path assignments in C2S and S2C communications. In S2C, the server broadcasts model weights to all clients simultaneously, whereas in C2S, each client uploads its updated model independently after completing local training. Despite this directional difference, both follow the same general scheduling process. In synchronous federated learning, the server must wait for all clients to finish uploading before proceeding to the next round~\cite{design-synchronous}. The overall workflow of the \emph{Flow Scheduler} is defined in Algorithm~\ref{alg:flow_scheduler}.

At the start of each training round, the scheduler enqueues clients to the Phase 1 queue ($Q_{p1}$) from the waiting queue ($Q_{wait}$), initiating \textit{Phase 1}. In this phase, it retrieves each client’s precomputed K-shortest paths from the \emph{Client Store} and selects the optimal path using a user-defined path selection strategy ($PSS$). Upon assignment, clients are enqueued to the Phase 2 queue ($Q_{p2}$). The scheduler then schedules \textit{Phase 2} to execute asynchronously after a configurable delay $S$ (default: five seconds), and returns to its main loop to monitor $Q_{wait}$ for newly arrived flows. Once the delay elapses, \textit{Phase 2} is triggered to process clients in $Q_{p2}$. In this phase, the scheduler updates client progress using metrics from the \emph{Client Store} and estimates completion times based on current throughput. Clients that complete their transfers are marked \emph{Done}, while those still in progress may re-enter $Q_{p1}$ for potential path reassignment if improved routes become available. Otherwise, they continue to \textit{Phase 2} until their transmissions conclude.

We implemented two $PSS$ variants, \emph{Constraint Optimization} and \emph{Greedy}, to showcase the flexibility of the framework. Although these serve as concrete examples, the system is designed to support any decision-making model. For instance, reinforcement learning can be integrated directly into the scheduling process without requiring any modifications to the core system.

\begin{algorithm}[t]
\caption{Flow Scheduler Algorithm}
\label{alg:flow_scheduler}
\begin{algorithmic}[1]
\State \textbf{Input:} Clients $C_{\text{all}}$, Path-Select Strategy $PSS$, Delay $S$
\State \textbf{Init:} $C_{\text{done}} \gets 0$, $C_{\text{total}} \gets |C_{\text{all}}|$
\State Queues: $Q_{\text{wait}} \gets \emptyset$, $Q_{\text{P1}} \gets \emptyset$, $Q_{\text{P2}} \gets \emptyset$

\Procedure{MainFlow}{}
    \While{$C_{\text{done}} < C_{\text{total}}$}
        \If{$Q_{\text{wait}} \ne \emptyset$}
            \State $Q_{\text{P1}} \gets Q_{\text{P1}} \cup Q_{\text{wait}}$
            \State $Q_{\text{wait}} \gets \emptyset$
        \EndIf
        \If{$Q_{\text{P1}} \ne \emptyset$}
            \State \Call{Phase1}{$Q_{\text{P1}}, M$}
        \EndIf
        \State \textit{/* Wait for next trigger or client updates */}
    \EndWhile
\EndProcedure

\Procedure{Phase1}{$Q_{\text{P1}}, M$}
    \State $P_{\text{map}} \gets \Call{PSS.ComputePaths}{Q_{\text{P1}}}$
    \State \Call{PSS.SwitchPaths}{$P_{\text{map}}$}
    \State $Q_{\text{P2}} \gets Q_{\text{P2}} \cup Q_{\text{P1}}$
    \State $Q_{\text{P1}} \gets \emptyset$
    \State \textit{/* Schedule Phase2 to run after interval $S$ */}
    \State \Call{SchedulePhase2}{$Q_{\text{P2}}, S$}
\EndProcedure

\Procedure{Phase2}{$Q$}
    \ForAll{$c \in Q$}
        \State \textit{/* Update remaining data from stats store */}
        \State $c.dataRemaining \gets \Call{ReadStats}{c}$
        \If{$c.dataRemaining \le 0$}
            \State $C_{\text{done}} \gets C_{\text{done}} + 1$
        \Else
            \State $Q_{\text{P1}} \gets Q_{\text{P1}} \cup \{c\}$
        \EndIf
    \EndFor
    \State $Q_{\text{P2}} \gets \emptyset$
\EndProcedure
\end{algorithmic}
\end{algorithm}

%% file: path_select.tex
\subsection{Constraint Optimization Model}
In this model, path assignment for weight communication is formulated as an optimization problem, aiming to minimize the longest completion time among all clients receiving model weights from the FL server. This objective is particularly important in synchronous FL, where the next training round cannot begin until every client completes the weight exchange. 

High latency increases RTT, which slows the growth of TCP’s congestion window, while packet loss leads to retransmissions and reduced throughput due to window backoff mechanisms~\cite{tcp_rfc5681}. These effects become more pronounced when multiple flows compete over congested links, reducing per-flow bandwidth and further extending transfer times. Effective path selection, therefore, must consider not only available capacity but also the compounded impact of latency and loss on performance.

The constraint optimization model accounts for these factors explicitly, allowing the scheduler to assign paths that minimize bottlenecks and maintain consistency across the network. For each client, it retrieves candidate paths from the \emph{Client Store} and gathers link-specific metrics from the \emph{Link Store}, ensuring that decisions are guided by both topology and real-time conditions.

For every candidate path $p$, defined as a sequence of directed links $\mathcal{L}_p$, we compute the effective RTT by summing the one-way delay of each link in both the forward direction for data transmission and the reverse direction for acknowledgments:
\begin{equation}
\text{RTT}_p = \sum_{l \in \mathcal{L}_p} \left( \text{lat}_{l, \text{forward}} + \text{lat}_{l, \text{reverse}} \right).
\end{equation}
The end-to-end packet loss for path $p$ is calculated under the assumption of independent link failures:
\begin{equation}
\text{PacketLoss}_p = 1 - \prod_{l \in \mathcal{L}_p} (1 - \text{ploss}_l),
\end{equation}
where $\text{ploss}_l$ denotes the packet loss probability of link $l$.

TCP Cubic has been the default congestion control algorithm in the Linux kernel since version 2.6.19~\cite{linux_cubic_default}. It is also the default in Windows and Apple's operating systems~\cite{rfc9438}. Compared to TCP Reno~\cite{ha2008cubic}, Cubic reacts less aggressively to packet loss, yet throughput still degrades when losses occur. To approximate this effect, we adjust RTT by scaling it with the square root of the packet loss rate, providing a tractable estimation of its impact on flow completion time~\cite{adj_rtt}:
\begin{equation}
\text{AdjRTT}_p = \text{RTT}_p \cdot \sqrt{ \text{PacketLoss}_p + \epsilon },
\end{equation}
where $\epsilon$ is a small constant for numerical stability.

To model path assignments, we define binary decision variables $x_{c,p} \in \{0,1\}$, indicating whether path $p$ is assigned to client $c$. Each client must select exactly one path:
\begin{equation}
\sum_{p \in \mathcal{P}_c} x_{c,p} = 1, \quad \forall c.
\end{equation}
To model link utilization, we introduce the variable $\text{ActiveFlows}_l$, representing the number of active client flows traversing link $l$:
\begin{equation}
\text{ActiveFlows}_l = \sum_{c} \sum_{p \in \mathcal{P}_c : l \in \mathcal{L}_p} x_{c,p}.
\end{equation}
Given $\text{cap}_l$ as the estimated available capacity of link $l$, the per-flow capacity along link $l$ becomes:
\begin{equation}
\text{FlowCap}_{l,p} = \frac{ \text{cap}_l }{ \text{ActiveFlows}_l }.
\end{equation}
A client's throughput on path $p$ is determined by the bottleneck link along that path, considering both capacity and latency penalties:
\begin{equation}
\text{Throughput}_{c,p} = \min_{l \in \mathcal{L}_p} \left( \frac{ \text{cap}_l }{ \text{AdjRTT}_p } \right).\label{eq:throughput}
\end{equation}

The remaining data volume to be transmitted for each client $c$ is denoted as $D_c$, which may vary across clients. The completion time for client $c$ using path $p$ is:
\begin{equation}
\text{CompletionTime}_{c,p} = \frac{ D_c }{ \text{Throughput}_{c,p} }.
\end{equation}

To model the worst-case scenario, we define a bottleneck variable $T$ as the maximum completion time over all clients and their selected paths:
\begin{equation}
T = \max_{c, p \in \mathcal{P}_c} \left( \frac{ D_c }{ \text{Throughput}_{c,p} } \cdot x_{c,p} \right).
\end{equation}
The optimization objective is to minimize the maximum completion time:
\begin{equation}
\min T.
\end{equation}

Using this model, \emph{ComputePaths} processes all clients concurrently, generating a new map that assigns each client to its optimal path based on the available candidates. \name then installs the corresponding flow rules on the network switches along the selected routes. To maintain stability, \emph{SwitchPaths} updates a client’s path only if two conditions are met: the recommended path differs from the current one, and at least two \openflow statistics polling intervals have elapsed since the last change. This conservative strategy reduces unnecessary switching, preserves measurement accuracy, and prevents disruptive oscillations in the network.

We implement this model using Google's CP-SAT solver~\cite{cpsatlp}, a constraint programming (CP) solver based on satisfiability (SAT) techniques. It exhaustively searches all feasible solutions, guaranteeing globally optimal results under defined constraints. Our implementation uses the OR-Tools Java API, version 9.12.

\subsection{Greedy Model}
Although the CP model offers globally optimal solutions, its computational cost scales poorly with network size, which can limit its practicality in larger deployments. To address this, we implement a \emph{Greedy PSS}, a more efficient alternative that makes locally optimal decisions at each step. While it does not ensure a global optimum, it delivers solid approximations with significantly lower overhead, making it well-suited for federated learning environments. 

For each client \( c \), the algorithm considers a set of \( K \)-shortest candidate paths \( \mathcal{P}_c = \{ p_1, p_2, \dots, p_K \} \). Each path \( p \in \mathcal{P}_c \) is assigned a score \( S(p) \), reflecting its expected performance based on available capacity, active flows, and latency penalties, computed similarly to the throughput formulation in equation~\eqref{eq:throughput}.

To prioritize clients effectively, the algorithm first computes the best available path score for each client:
\[
S_{\text{max}}(c) = \max_{p \in \mathcal{P}_c} S(p).
\]
The clients are then sorted in ascending order of their \( S_{\text{max}}(c) \) values:
\[
c_1, c_2, \dots, c_n \quad \text{such that} \quad S_{\text{max}}(c_1) \leq \dots \leq S_{\text{max}}(c_n).
\]
This sorting ensures that clients with weaker path options are prioritized, securing the best possible paths before network resources become saturated.

For each candidate path, we normalize the scores using min-max normalization to ensure fair comparisons across different metrics:
\[
x' = \frac{x - x_{\text{min}}}{x_{\text{max}} - x_{\text{min}}},
\]
where:
\[
x_{\text{min}} = \min_{p \in \mathcal{P}_c} S(p), \quad x_{\text{max}} = \max_{p \in \mathcal{P}_c} S(p).
\]

In \emph{ComputePaths}, the model processes clients one at a time, selecting for each the candidate path with the highest normalized score. \name then installs the corresponding flow rules along the selected path in the network. To prevent instability, \emph{SwitchPaths} triggers only when the new path offers at least a 30\% improvement in score and two \openflow polling intervals have passed since the last change.



\begin{table*}[t]
  \centering
\caption{Comprehensive comparison of the \rfwd baseline, \freecap, and the proposed solutions \cp and \greedy across three network setups. Metrics are divided into two categories, training-time metrics and network-stability metrics. Lower values indicate better performance, reflecting improved training speed and enhanced network stability.}
\vspace{-2mm}
  \renewcommand{\arraystretch}{1.25} 
  \setlength{\tabcolsep}{2pt} 
  \begin{tabular}{l|cccc||cccc||cccc}
    \toprule
    \textbf{Metric} & \multicolumn{4}{c||}{\textbf{Topology $E1$}} & \multicolumn{4}{c||}{\textbf{Topology $E2$}} & \multicolumn{4}{c}{\textbf{Topology $E3$}} \\
     & \multicolumn{4}{c||}{(15 switches, 15 clients)} & \multicolumn{4}{c||}{(30 switches, 25 clients)} & \multicolumn{4}{c}{(50 switches, 50 clients)} \\
    \cmidrule(lr){2-5} \cmidrule(lr){6-9} \cmidrule(lr){10-13}
    & \textbf{$RFWD$} & \textbf{$FreeCap$} & \textbf{$SF_{Greedy}$} & \textbf{$SF_{CP}$} 
    & \textbf{$RFWD$} & \textbf{$FreeCap$} & \textbf{$SF_{Greedy}$} & \textbf{$SF_{CP}$} 
    & \textbf{$RFWD$} & \textbf{$FreeCap$} & \textbf{$SF_{Greedy}$} & \textbf{$SF_{CP}$} \\
    \midrule
    \textbf{Time to 60\% Acc (min)}   & 12.4 & 11.7 & 8.2  & \textbf{7.7}    & 20.0 & 19.7 & \textbf{10.5} & 10.7   & 42.2 & 29.7  & \textbf{20.0} & 21.0 \\
    \textbf{Time to 80\% Acc (min)}   & 37.2 & 33.5 & 22   & \textbf{21}     & 56.0 & 50.7 & \textbf{29.6} & 32.1   & 118.3 & 89.5 & 67.7 & \textbf{65} \\
    \textbf{Avg Round Time (sec)}     & 53.1 & 46.8 & 32.9 & \textbf{32.1}   & 67 & 60.8   & \textbf{35.6} & 38.5   & 94.6 & 71.6  & 54   & \textbf{52} \\
    \textbf{S2C Path Reassignments} & 0 & 0 & 0 & 0 & \textbf{0} & 0 & 12 & 10 & \textbf{0} & 0   & 85 & 70 \\
    \textbf{C2S Path Reassignments} & 0 & 0 & 0 & 0 & \textbf{0} & 0 & 30 & 23 & \textbf{0} & 0   & 119 & 90 \\
    \textbf{gRPC Timeouts}       & 83 & 42 & 12 & \textbf{7} & 129 & 94 & \textbf{9} & 15 & 797 & 651 & 150   & \textbf{143} \\
    \bottomrule
  \end{tabular}
  \label{table:all_results}
  \vspace{-2mm}
\end{table*}

%% file: evaluations.tex
We evaluate two variations of \name that explore different flow scheduling strategies. The first, \cp, uses the CP model for  S2C communication to enable globally coordinated updates, and the Greedy model for C2S communication to match the clients' asynchronous behavior. The second, \greedy, applies the Greedy model to both S2C and C2S, offering a fully lightweight option with minimal scheduling overhead. We benchmark both \cp and \greedy against the reactive forwarding method (\rfwd), which uses the built-in shortest-path algorithm of ONOS, and \freecap~\cite{mahmod_freecap}, which forwards traffic through the path with the freest bottleneck link capacity.

\subsection{Testbed}
To evaluate our approach, we built a modular testbed that emulates diverse network scenarios with varying complexity, topology, traffic patterns, and numbers of FL clients. It comprises three components, Topology Creator, Congestion Simulator, and Experiment Runner, coordinated by a central module for streamlined configuration. The testbed supports customizable topologies via the TopoHub repository, emulates congestion with containerized iperf3 flows, and tracks real-time training and network metrics through an automated FL pipeline. For more details about the testbed, check~\cite{testbed-paper}.
\subsection{Experiments}
We evaluated our solution on Gabriel network topologies of varying complexity, client count, and traffic. We used three topologies from Topohub~\cite{intro13-topohub}: $E1$ (15 nodes, 9 variations, 15 FL clients), $E2$ (30 nodes, 7 variations, 25 clients), and $E3$ (50 nodes, 8 variations, 50 clients). Federated training used Flower AI~\cite{intro12-flower} with PyTorch~\cite{pytorch}, exchanging model updates over gRPC. To simulate congestion, we deployed iperf3 servers and clients in containers at both ends of each link, generating sequential TCP flows based on a Poisson distribution $X \sim \text{Poisson}(\lambda)$, where $\lambda$ reflects the ECMP utilization from the Topohub configs file.

We trained on the CIFAR-10 dataset~\cite{cifar10} with centralized evaluation at the server. We deliberately used a single model and dataset, as the model size, being the dominant factor in data exchange per round, has a greater impact on network performance than the specific architecture or dataset. Training spanned 40 rounds for $E1$, 50 for $E2$, and 75 for $E3$, with one local epoch per round. To ensure fair comparison, the number of stored K-shortest paths in the \emph{Client Store} was capped at 10, 15, and 25 for $E1$, $E2$, and $E3$, respectively.

\tablename~\ref{table:all_results} summarizes the performance across evaluated metrics, grouped into two categories: (1) \emph{training-time metrics}, including the time to reach 60\% and 80\% accuracy and average round time, and (2) \emph{network-stability metrics}, including S2C and C2S path reassignments and gRPC timeouts. Path reassignments measure the frequency of route updates, impacting network stability if excessive. Similarly, gRPC timeouts, arising from request delays, degrade training efficiency and system reliability if frequent.

In topology $E1$, \name significantly outperformed \rfwd and \freecap in training-time metrics. \cp reached 80\% accuracy in 21 minutes, representing a 43\% improvement over \rfwd and 37\% over \freecap. Average round time dropped to 32 seconds, compared to 53 seconds for \rfwd and 46.8 seconds for \freecap. Compared to \greedy, \cp achieved a modest improvement of approximately 5\%. Network stability also improved significantly, with only 7 gRPC timeouts, an 83\% reduction relative to \freecap.

In the more complex topology $E2$, the advantages became clearer. \greedy reached 80\% accuracy in 29.6 minutes, representing a notable reduction of 47\% and 41.5\% compared to \rfwd and \freecap, respectively. Average round time decreased notably to 35 seconds, down from 67 seconds with \rfwd and 60 seconds with \freecap. Network stability remained robust, with fewer path reassignments, 42 for \greedy and 33 for \cp, and substantially fewer gRPC timeouts: 9 for \greedy and 15 for \cp, compared to 94 for \freecap and 129 for \rfwd.

With topology $E3$ nearly double the size of $E2$, complexity grew significantly. Nevertheless, \cp reached the target accuracy in 65 minutes, reducing training time by 45\% compared to \rfwd and by 27\% compared to \freecap. The average round time also improved notably, dropping to 52 seconds for \cp and 54 seconds for \greedy, whereas \freecap required 71 seconds per round. Although the larger scale led to more frequent path reassignments, network reliability improved substantially. The number of gRPC timeouts dropped to 143, representing an 82\% reduction relative to \rfwd.

While \cp consistently outperformed the baselines, its advantage over \greedy depended on network size and dynamics. In $E1$, the small network let \cp compute optimal paths quickly, giving it a slight edge. In $E2$, the increased dynamics favored \greedy, whose fast, low-overhead updates outpaced \cp’s slower route computations. In $E3$, \cp’s global optimization had a stronger impact, as the benefits of better paths outweighed its initial delays. Overall, \cp is better for large, complex networks where path quality matters, while \greedy is better for fast-changing or resource-limited environments.

\begin{figure}[tbp]
	\begin{center}
	\includegraphics[width=0.90\linewidth]{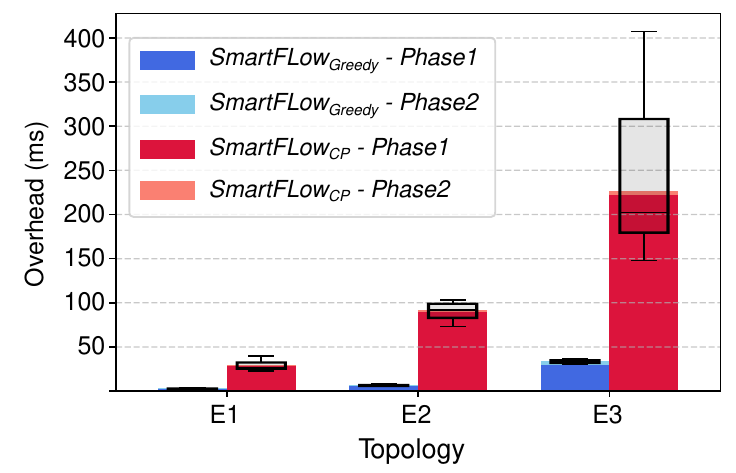}
		\vspace{-1mm}
		\caption{Computational overhead analysis during Phase 1 and Phase 2 processing across different network topologies.}
		\label{fig:overhead_all}
	\end{center}
    \vspace{-6mm}
\end{figure}

\figurename~\ref{fig:overhead_all} compares the computational overhead of \greedy and \cp during Phases 1 and 2 across topologies $E1$–$E3$. In $E1$, \greedy stayed near 10 ms, while \cp was higher at ~35 ms. In $E2$, overhead grew to ~15 ms for \greedy and ~90 ms for \cp, with spikes near 100 ms. Under the most demanding $E3$, \greedy reached ~40 ms, whereas \cp averaged 200 ms and peaked at 400 ms. These results underscore \cp’s scalability issues and \greedy’s ability to maintain low overhead.

%% file: conclusion.tex
In this paper, we address communication inefficiencies in cross-silo FL, mainly caused by network congestion and client stragglers, which limit training performance and system scalability. To tackle this, we propose \name, an SDN-integrated framework that dynamically optimizes client-server communication paths based on real-time network conditions. By carefully balancing network utilization and stability, \name reduces communication time by up to 47\% compared to existing routing methods, with minimal overhead. This efficiency makes it suitable for practical deployment in large-scale and heterogeneous FL topologies, overcoming critical bottlenecks in real-world systems.